\documentclass{aa}
 
\usepackage{graphics}
\usepackage{astron}

\begin{document}

\thesaurus{11(11.09.1;11.11.1;11.19.6)}

\titlerunning{Vertical extent and kinematics of HI in NGC~2403} 
\title{The vertical extent and kinematics of the HI in NGC~2403} 

\author{W.E.~Schaap \inst{1}  
  \and R.~Sancisi \inst{2,1}
  \and R.A.~Swaters \inst{3}}

\institute{Kapteyn Astronomical Institute, University of Groningen,
  P.O.~Box 800, 9700 AV Groningen, The Netherlands 
  \and Osservatorio Astronomico, Bologna, Italy
  \and Department of Terrestrial Magnetism, Carnegie Institution of
  Washington, Washington, USA}

\offprints{W.E.~Schaap}
\mail{wschaap@astro.rug.nl}

\date{Received ~~~~~~~~~~~~~~~~~~~~~~~ / Accepted }

\maketitle

\begin{abstract}
  
  The neutral hydrogen line profiles along the major axis of the
  nearby spiral galaxy NGC~2403 show a wing towards the systemic
  velocity.  This asymmetry can be explained with the presence of an
  abnormally thick HI disk (FWHM $\sim$ 5 kpc) or with a two-component
  structure: a thin disk and a slowly rotating, thicker (1-3 kpc) HI
  layer. The latter model gives a better representation of the
  observations. These results throw a new light on the disk-halo
  connection in spiral galaxies. In particular, the decrease of
  rotational velocity with height above the plane may be the result of
  a galactic fountain flow. A vertically extended, slowly rotating HI
  layer may be common among spiral galaxies with high levels of star
  formation.
  

\end{abstract}

\section{Introduction}

In recent years there has been a growing interest for the vertical
structure of the HI disks of spiral galaxies and for the disk-halo
connection.  Evidence for HI gas flows between disk and halo comes
from the detection of large vertical motions in several galaxies
viewed close to face-on (Dickey et al.~1990; Kamphuis 1993; Schulman
\& Bregman 1994) and the connection of this high velocity gas with HI
holes and star formation activity in the disk (Kamphuis \& Sancisi
1993). More evidence comes from the study of edge-on galaxies like
NGC~891 in which the HI has been found to extend up to at least 5 kpc
into the halo, where it seems to rotate about 25 km ${\rm s}^{-1}$
more slowly than in the plane (Swaters et al.~1997). The combination
of the results from face-on and edge-on galaxies leads to the picture
of effervescent galaxies (Sancisi et al.~1996), consistent with
galactic fountains models (Bregman 1980; Spitzer 1990).

\begin{figure*}[t]
  \center{\scalebox{.8}{\includegraphics{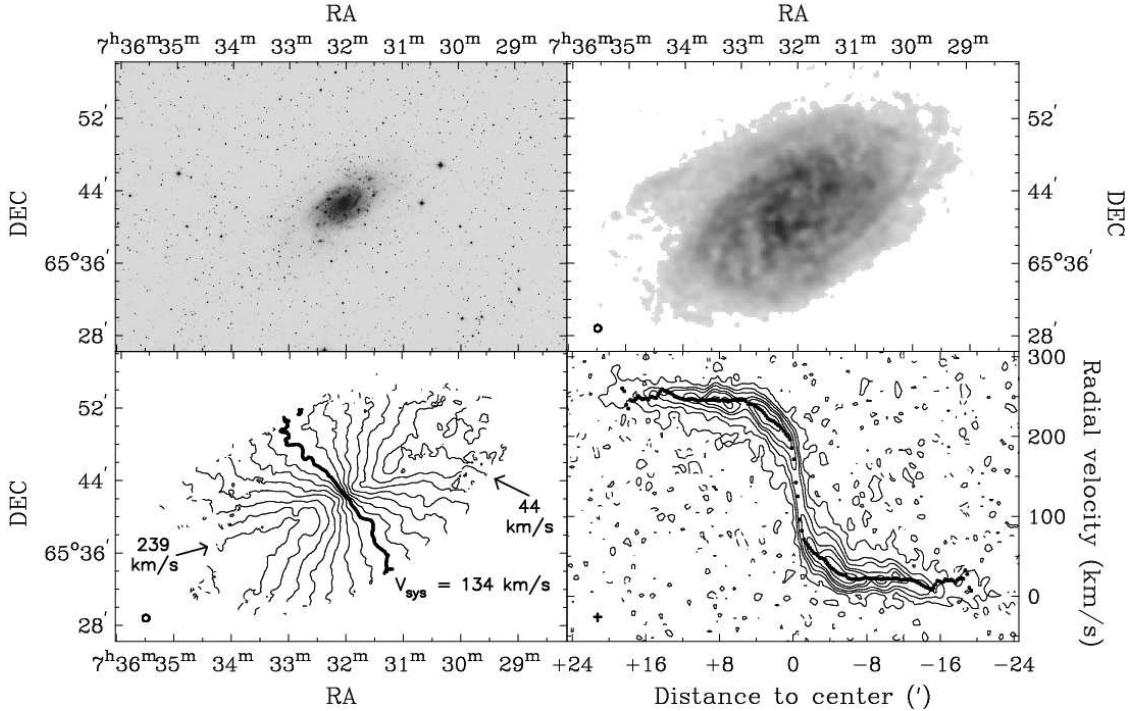}}}
  \parbox{16.5cm}{\caption{\small Optical image (top left; from the
      Palomar Digitized Sky Survey), HI column density map (top right;
      the column densities range from $7.5 \cdot 10^{19}$ to $3.3
      \cdot 10^{21}$ atoms/cm$^2$) and HI velocity field (bottom left;
      the contours run from 14 to 254 km ${\rm s}^{-1}$ in steps
      of 15 km ${\rm s}^{-1}$.).  The beam ($45'' \times 45''$) is
      shown in the lower left corner.  The bottom right panel shows
      the HI position-velocity map along the major axis
      (PA=$124^{\circ}$).  Contours are -2.5 (dashed), 2.5
      (1.8$\sigma$), 10, 20, 40, 60 and 100 mJy/beam.  The black dots
      indicate the rotation curve derived by Sicking (1997).  The
      angular and velocity resolutions ($45'', 8.2$ km ${\rm s}^{-1}$)
      are indicated by the cross in the lower left corner.
\label{f: n2403panel}}}\end{figure*}

The galaxy studied here, NGC~2403, has an intermediate inclination
(i=$61^{\circ}$).  The consequence is that the measured line-of-sight
velocities are a combination of rotational, radial and vertical
motions and the column densities are integrated along an oblique
line-of-sight.  Therefore, the interpretation is less straightforward
than in face-on or edge-on galaxies. However, there is the advantage that
information is obtained on both the vertical density structure and the
vertical kinematics of the HI for the same object.

For this study we have used the HI observations obtained by Sicking
(1997) with the Westerbork Radio Telescope, which have about a factor
of 2 better sensitivity and a higher velocity resolution than those of
Wevers et al.~(1986). Fig.~1 shows the optical image, the total HI
density distribution, the velocity field and the HI position-velocity
map along the major axis.  The latter shows that the HI line profiles
at any position along the major axis are not symmetrical with respect
to the rotation velocities, as they would be if they were determined
by random motions only. Instead, they are systematically more extended
towards the systemic velocity. This striking asymmetry is particularly
obvious in the lowest contour and extends systematically over almost
the whole major axis. It was already noticeable in the maps produced
by Wevers et al.~(1986) and by Begeman (1987) and the puzzle presented
by those early observations has motivated this study. The presence of
such a `beard' is remarkable considering the size of the beam (see
lower left in fig.~1) with respect to the size of the galaxy. Usually
such an asymmetry is seen in edge-on galaxies or in galaxies which are
not well resolved. In such cases the telescope beam `sees' not only
the emission from a small area on the major axis but also larger areas
away from it which have lower line-of-sight velocities and therefore
cause the observed asymmetry.  NGC~2403 is neither highly inclined nor
poorly resolved.

What is the origin of the `beard'? It is clear that it cannot simply
be explained by gas moving perpendicularly away from or towards the
disk, because that would produce extensions symmetric with respect to
the rotation velocity. Neither can it be the result of deviations from
axial symmetry or circular motion. These would affect the kinematics
of the disk and produce visible effects in the velocity field, but not
a low density asymmetry as observed.  A likely explanation for the
observed asymmetry is that not all of the HI is concentrated in a thin
disk, but that part of it is in a vertically extended component. In
this case, given the inclination of 61 degrees, a line-of-sight to a
point on the major axis will also intercept the HI located above and
below the plane which has line-of-sight velocities lower than those in
the plane. This will produce a systematic broadening of the HI
profiles towards the systemic velocity of the kind seen in fig.~1. It
is this effect that we study here with 3-D models of the density
distribution and kinematics of the HI in NGC~2403.

\begin{figure*}
  \center{\scalebox{.85}{\includegraphics{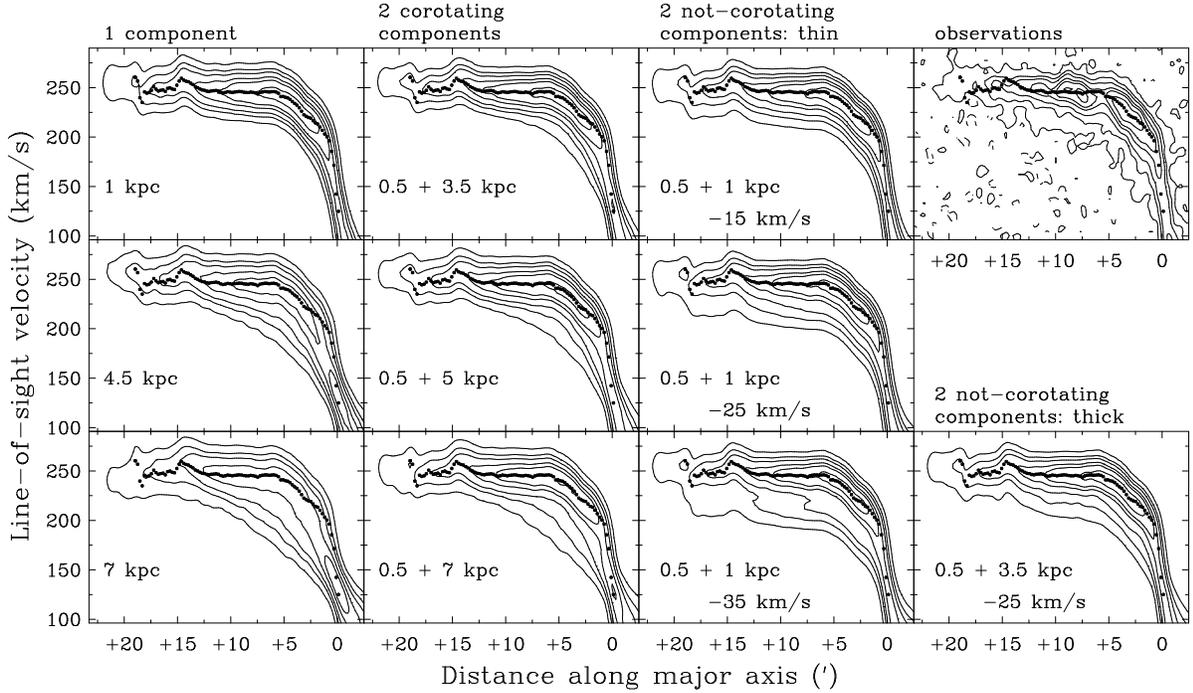}}}
  \parbox{16.5cm}{\caption{\small Position-velocity maps (receding
      side, see fig.~1) along the major axis of NGC~2403
      (PA=$124^{\circ}$) of the observations (top right panel, contour
      levels are -2.5, 2.5 (1.8$\sigma$), 10, 20, 40, 60 and 100
      mJy/beam) and of different models. The types of models are
      indicated at the top of each column. The FWHM values for the
      thin and thick components and their velocity differences are
      given in the panels. The angular and velocity resolutions are
      $45''$ and $8.2$ km ${\rm s}^{-1}$.}}
\label{f: lvmodels}
\end{figure*}
\section{Modelling}

We consider the possibilities of both a thick HI disk (as opposed to
the usually assumed thin layer), and of a two-component structure,
with a thin layer and a thicker but less dense one. This thick layer
may be corotating with the thin one or may be rotating more slowly.
Both cases are examined. The reason for considering the slower
rotation is that a non-corotating disk and halo have been proposed by
Swaters et al.~(1997) as the most likely interpretation for the HI
observations of the edge-on galaxy NGC~891.

Axial symmetry and circular motions are assumed throughout. This
assumption is supported by the results of Schoenmakers et al.~(1997)
who found from their harmonic analysis of the HI velocity field of
NGC~2403 that non-circular motions are not important in this galaxy.
Therefore, we have modelled NGC~2403 with a set of concentric rings,
each ring characterized by a radius, a circular velocity, a velocity
dispersion and a column density. Centre, systemic velocity,
inclination and position angle, scaleheight and vertical density
profile were chosen to be the same for all rings. The values adopted
here are, unless specified otherwise, those derived by Sicking (1997).

The thin component always has a FWHM thickness of 0.5 kpc, the
thickness of the thick layer is a free parameter. For the vertical
density profile we have assumed a Gaussian distribution. We have also
tried an exponential distribution, but the results do not seem to
change significantly. In the case of non-corotation the circular
velocity of the thicker component has been lowered by a fixed amount,
for example 25 km ${\rm s}^{-1}$, in the flat part of the curve and a
proportionally smaller one in the inner parts. For the velocity
dispersion we have used a value of 7 km ${\rm s}^{-1}$ as derived for
the profile widths in the regions of the flat part of the rotation
curve.  We have also constructed models with higher velocity
dispersions for the thick components, up to 30 km ${\rm s}^{-1}$, but
the results do not differ significantly.

The two-component models have been constructed by adding two
one-component models. This is obviously an oversimplification as it is
more likely that any vertical decrease of density and velocity would
be gradual.

\section{Results}

The model data cubes have been inspected and analyzed in the same way
as the observed cubes. For the comparison of the models and the
observations we have used the channel maps and the position-velocity
maps along the major and minor axes. Here we present only the
position-velocity maps along the major axis and only a small number of
models, but we have explored the whole relevant parameter space.

Initially, we tried a model with a one-component gas layer. Cuts along
the major axis for various thicknesses are shown in the first column
of fig.~2. The observations are in the top right panel. It is
immediately clear that a thin disk model does not reproduce the
observed asymmetry. A thickness of about 5 kpc (FWHM) would be needed
to explain it. It is obvious that such a layer would be
unrealistically thick, much thicker than that of our Galaxy, which
according to Dickey and Lockman (1990) has a FWHM of only 220 pc, or
that of NGC~891, for which most of the HI is in an unresolved layer
with a FWHM thickness of less than 1 kpc (Swaters et al.~1997).

Therefore, we explored the possibility of having, instead of one very
thick Gaussian disk, a relatively thin disk of high density and in
addition a vertically more extended but lower density layer. For this
we have built a model with two components, one thin and one thick,
both rotating at the same velocity.  The FWHM of the thick component
and the column density ratio between the two components are free
parameters.  We found, however, that there is little freedom in the
choice of this ratio and we fixed it tentatively at 1:1. With a lower
density for the thick component only a weak asymmetry would appear.
The best model seems the one with a thick component of FWHM close to 5
kpc (see fig.~2). Hence this model is not significantly different from
the one-component model and is equally unrealistic as it requires
unreasonable amounts of HI at large distances from the plane.

Finally, we released the condition of corotation and explored the
effects of a decrease of the rotational velocity with distance to the
plane. For this we have constructed models with a slowly rotating
thick component. Again, the FWHM of the thick component and the column
density ratio are free parameters. But here as an additional parameter
we have the velocity decrease. A thin-to-thick disk density ratio of
4:1 was found to give the best comparison with the data.  With a
higher ratio the asymmetry would tend to disappear, whereas a lower
ratio would not reproduce the observed density contrast between the
profile peak and its wings. For the thick component a FWHM of 1 kpc
was adopted. For this thickness the best agreement with the
observations is obtained with the 25 km ${\rm s}^{-1}$ decrease (see
fig.~2).  Models with a thick component of larger FWHM than 1 kpc were
also tried. For example, in the lower right of fig.~2, we show a model
that also reproduces the observed asymmetry well, and in which the
thick component, still with a 25 km ${\rm s}^{-1}$ decrease in
rotation velocity, has a thickness of 3.5 kpc. For this thickness the
optimal column density ratio of the thin and thick disk is 6:1. This
illustrates that it is difficult to distinguish between an extended,
low density and a less extended, higher density vertical distribution.

\section{Discussion and conclusion}

From section 3 it is clear that all three types of models can explain
the observed asymmetry to some extent. However, the first two models
require unrealistic amounts of HI above the plane. The third model has
a much more realistic vertical distribution, with less than 20\% of
the gas in the thick component, and it requires that the gas above the
plane rotates more slowly than the gas in the disk, by about 25 km
${\rm s}^{-1}$.  This model also appears to reproduce the observed
position-velocity diagram better than the other models, in particular
near the center, where the HI has a narrow and peaked distribution,
and at large radii, where low level wings are visible.

Note that in the observed position-velocity diagram (fig.~1) the wings
near the center extend almost to the systemic velocity. Clearly, such
features cannot be reproduced in models with corotation, and not even
in those with a velocity decrease as high as 35 km ${\rm s}^{-1}$. To
explain these extended wings a much larger velocity decrease is
needed, probably at least 50 km ${\rm s}^{-1}$.

What produces such a thick HI layer and what causes its slower
rotation?  Galactic fountain models (Bregman 1980; Spitzer 1990) may
provide the framework for an answer.  The fountain is formed by hot
gas rising from the disk, its energy derived from stellar winds and
supernova explosions.  In the halo region the gas cools and condenses
into clouds that fall back onto the disk.  As the gas moves up, the
centrally directed gravitational force decreases, and the gas moves
outwards.  Due to conservation of angular momentum the azimuthal
velocity decreases.  This decrease will be most pronounced in the
central parts of the galaxy, where a fixed radial displacement will
result in a larger velocity decrease than in the outer regions.  Using
thermohydrodynamic models, Struck \& Smith (1999) have recently shown
that a reduced circular velocity above and below the plane is expected
to be present in turbulent disks as a result of radial motions driven
by star formation activity.

NGC~2403 appears to have sufficient star formation activity to drive a
galactic fountain.  This is for instance indicated by its large number
of HII regions (Sivan et al.~1990).  Four of these are exceptionally
bright, comparable to the most massive starburst region in the Local
Group, the 30 Doradus complex (Drissen et al.~1999). Furthermore,
Thilker et al.~(1998) have found that the surface of NGC~2403 is
covered by shells and fragmentary structures which are likely to have
formed as a result of star formation activity.  They have also found
that these structures are part of a diffuse component of neutral
hydrogen extending at least 400 pc from the plane.

The overall picture of NGC~2403, as suggested by the modelling, is
reminiscent of that of NGC~891. For this galaxy it was found (Swaters
et al.  1997) that the thick component has a FWHM of about 4 kpc, and
this gas appears to rotate more slowly than the gas in the disk by
about 25 km ${\rm s}^{-1}$. In the central parts the velocity decrease
was found to be much larger, perhaps up to 100 km ${\rm s}^{-1}$.
 
In conclusion, we have presented evidence that the thin hydrogen disk
of NGC~2403 is surrounded by a vertically extended layer of HI, which
rotates more slowly than the disk. The observational picture is very
similar to that found for NGC~891. These results suggest that a
vertically extended, slowly rotating HI layer may be common among
spiral galaxies, at least among those with high levels of star
formation. Such extended HI layers have, however, very low surface
densities and therefore very sensitive observations are needed to
detect them.

\section{Acknowledgements}

We thank F.J.~Sicking for kindly making available the data and
U.J.~Schwarz for helpful discussions.

\end{document}